\documentclass [a4paper,fleqn, 12pt]{article}
\usepackage{graphicx}
\usepackage[small]{subfigure,epsfig}

\usepackage {amsmath} \usepackage{amssymb} \usepackage{cite}

\begin{document}

\title{Analytical properties and exact solutions of the Lotka--Volterra competition  system}

\author{Nikolay A. Kudryashov, \and Anastasia S. Zakharchenko}

\date{Department of Applied Mathematics, \\
National Research Nuclear University MEPhI \\
(Moscow Engineering Physics Institute), \\
31 Kashirskoe Shosse, 115409 Moscow, Russian Federation}


\maketitle

\begin{abstract}
The Lotka--Volterra competition  system with diffusion is considered. The Painlev\'e property of this  system is investigated. Exact traveling wave solutions of the Lotka--Volterra competition  system are found. Periodic solutions expressed in terms of the Weierstrass elliptic function are also given.

\end{abstract}


\section{Introduction}

%
%
%

In mathematical ecology, it has been proposed that systems of reaction-diffusion equations can describe the interaction of biological species which move by diffusion.

Denote by $U(X,T)$ and $V(X,T)$ population densities at position $X$ and time $T$ of two different species.
Assuming that they compete for the same food, we obtain the possible model. The model under consideration is the modified competition Lotka--Volterra system with diffusion~\cite{Murray2003,Okubo-Murray1989}
\begin{equation} \label{eq:1.1}
\begin{gathered}
\begin{split}
&\frac{\partial U}{\partial T}=D_1\,\frac{\partial^2 U}{\partial X^2}+a_1U(1-b_1U-c_1V),\\
&\frac{\partial V}{\partial T}=D_2\,\frac{\partial^2 V}{\partial X^2}+a_2V(1-b_2V-c_2U),
\end{split}
\end{gathered}
\end{equation}
where $a_i$ are net birth rates, $1/b_i$ are carrying capacities, $c_i$ are competition coefficients and $D_i$ are  diffusion coefficients. All of the above constants are non-negative. The interaction terms represent logistic growth with competition. We assume the first population outcompete the second so
\begin{equation} \label{eq:1.1cond}
b_2>c_1, \quad c_2>b_1.
\end{equation}

We nondimensionalise the system by setting
\begin{equation} \label{eq:1.2}
\begin{gathered}
x=X\left(\frac{a_1}{D_1}\right)^{1/2}, \quad t=a_1T, \quad u=b_1U, \quad v=b_2V, \\
\gamma_1=\frac{c_1}{b_2}, \quad \gamma_2=\frac{c_2}{b_1}, \quad d=\frac{D_2}{D_1}, \quad \alpha=\frac{a_1}{a_2}.
\end{gathered}
\end{equation}
and \eqref{eq:1.1} becomes
\begin{equation} \label{eq:1.3}
\begin{gathered}
\begin{split}
&\frac{\partial u}{\partial t}=\frac{\partial^2 u}{\partial x^2}+u(1-u-\gamma_1v),\\
&\frac{\partial v}{\partial t}=d\,\frac{\partial^2 v}{\partial x^2}+\alpha v(1-v-\gamma_2u).
\end{split}
\end{gathered}
\end{equation}
Because of \eqref{eq:1.1cond} we have
\begin{equation} \label{eq:1.3cond}
\gamma_1<1, \quad \gamma_2>1.
\end{equation}

Traveling wave exact solutions of system~\eqref{eq:1.1} have been studied comprehensively: existence, stability and uniqueness.
Hosono~\cite{Hosono1989} investigated the existence of traveling waves of~\eqref{eq:1.3} with~\eqref{eq:1.1cond} and with boundary conditions $u=1,\,v=0\, \mbox{at}\,  {z\rightarrow -\infty},\,u=0,\,v=1\, \mbox{at}\,  {z\rightarrow +\infty}$ under certain restrictions on the values of the parameters. He showed that in general case, the system of differential equations~\eqref{eq:1.3} cannot be solved analytically. Some analytical results can be obtained only in the special case where $d=\alpha=1,\, \gamma_1+\gamma_2=2$.
Gardner~\cite{Gardner1982} investigated the existence of traveling fronts connecting the equilibrium states $(1,0)$ and $(0,1)$ for the bistable case: $\gamma_1,\gamma_2>1$. In \cite{Kan-on1995}, the author obtained the monotone dependence of the wave speed on parameters for the bistable case. Guo and Lin \cite{GuoLin2013} investigated (for the bistable case too) the sign of the wave speed and its dependence on parameters, because the sign of the propagation speed determines which species become dominant.
The existence of traveling waves and the minimal wave speed for the monostable case ($0<\gamma_1<1<\gamma_2$ or $0<\gamma_2<1<\gamma_1$) are shown in~\cite{Hosono1989,Hosono1998},~\cite{GuoLiang2011},~\cite{Kan-on1997}. When time delays are incorporated in system~\eqref{eq:1.1}, Li et al.~\cite{Li2006}, Li~\cite{Li2008}, Huang and Zou~\cite{HuangZou2002} proved the existence of traveling fronts, see also \cite{YuYuan2011,LiLi2009} and references therein.
Bao and Wang~\cite{BaoWang2013} considered existence and stability of time periodic traveling waves in a periodic bistable Lotka--Volterra competition system.
There are also many papers on traveling wave solutions of lattice dynamics system arising in competition models \cite{GuoWu2011,GuoWu2012,GuoWu2012b}.
Besides traveling wave solutions some authors investigated the existence of entire solutions~\cite{GuoWu2010, Morita2009,WangLv2010}.

So, despite the large amount of works investigating traveling waves in system~\eqref{eq:1.1}, there are only several works where explicit exact solutions are found. Rodrigo and Mimura~\cite{RodrigoMimura2000} considered system~\eqref{eq:1.1} with $D_1\neq D_2$.  Applying the certain ansatz to the system, they obtained exact traveling wave and standing wave solutions under some parameter restrictions and for various correlations between $D_1$ and $D_2$. But solutions for the case of $D_1=D_2$ are not presented among the obtained solutions. It is interesting to investigate analytical properties and exact traveling wave solutions of system~\eqref{eq:1.1} for the case $D_1=D_2$ ($d=1$).

The aim of this work is to investigate analytical properties of system of equations~\eqref{eq:1.3}, particularly the Painlev\'e property, and to obtain some exact traveling wave solutions of this system in the case of $D_1=D_2$ ($d=1$).

\section{Painlev\'e analysis of the Lotka--Volterra competition system}
Let us investigate system~\eqref{eq:1.3} on the Painlev\'e property. This is an important analytical property, because it is well known that the Painlev\'e test for nonlinear differential equations is a powerful approach for testing integrable differential equation.

At the first step of investigation we need to reduce our system of partial differential equations to an ordinary differential equation system. Using traveling wave variables
\begin{equation} \label{eq:1.4}
\begin{gathered}
u(x,t)=y(z),\quad v(x,t)=w(z), \\
z=k\,x-C_0\,t \quad (k\,\neq0)
\end{gathered}
\end{equation}
we obtain
\begin{equation} \label{eq:1.3a}
\begin{gathered}
{k}^{2}{y_{zz}}+{C_0}\,y_z +y\left(1-y-\gamma_1w \right)=0,\\
{k}^{2}\,w_{zz}+{C_0}\,w_z +\alpha w \left(1-w-\gamma_2y\right)=0.
\end{gathered}
\end{equation}

For simplicity we reduce obtained system~\eqref{eq:1.3a} to a single equation in the form
\begin{equation} \label{eq:1.3b}
\begin{gathered}
\begin{split}
&w^5\alpha^2\gamma_1\gamma_2-w^4\alpha^2\gamma_1\gamma_2+2C_0w_zw^3\alpha-2C_0w_zw^2\alpha+2w_{zz}w^3\alpha k^2 \\
&-2w_{zz}w^2\alpha k^2
-C_0 w_z w^3 \alpha \gamma_1 \gamma_2-w_{zz} w^3 \alpha \gamma_1 \gamma_2 k^2-2 w_{zzz} w_z w \alpha \gamma_2 k^4 \\
&+2 C_0 w_{zzz} w^2 \alpha \gamma_2 k^2-w_{zz}^2 w k^4-w^4 \alpha^2 \gamma_2+w^3 \alpha^2 \gamma_2-C_0^2 w_z^2 w \\
&+w_{zz}^2 w \alpha \gamma_2 k^4-2 w_{zz} w_z^2 \alpha \gamma_2 k^4+w_{zz} w^3 \alpha^2 \gamma_2 k^2-w^2w_{zzzz}\alpha\gamma_2 k^4 \\
&+C_0^2w_{zz}w^2\alpha\gamma_2+w_{zz}w^2\alpha\gamma_2k^2-2C_0w_{zz}w_zwk^2+2C_0w_z^3\alpha\gamma_2k^2\\
&-C_0 w_z w^3 \alpha^2 \gamma_2-C_0^2 w_z^2 w \alpha \gamma_2+C_0 w_z w^2 \alpha \gamma_2-w^5 \alpha^2+2 w^4 \alpha^2\\
&-w^3 \alpha^2-4 C_0 w_{zz} w_z w \alpha \gamma_2 k^2=0
\end{split}
\end{gathered}
\end{equation}
and will investigate the Painlev\'e property of this equation.

The leading members corresponding to Eq.\eqref{eq:1.3b} have the form
\begin{equation} \label{eq:1.3c}
\begin{gathered}
\begin{split}
&w^5\alpha^2\gamma_1\gamma_2-w^5\alpha^2+w^2w_{zzzz}\alpha\gamma_2k^4-w_{zz}w^3\alpha^2\gamma_2k^2+2w_{zz}w_z^2\alpha\gamma_2k^4\\
&- w_{zz}^2w\alpha\gamma_2k^4-w_{zz}^2wk^4-2w_{zzz}w_zw\alpha\gamma_2k^4-w_{zz}w^3\alpha\gamma_1\gamma_2k^2\\
&+2w_{zz}w^3\alpha k^2=0.
\end{split}
\end{gathered}
\end{equation}

Substituting $w=a_0/z^{p}$ \cite{Ablowitz01, Ablowitz02} into equation \eqref{eq:1.3c} we get
\begin{equation} \label{eq:1.40}
\begin{gathered}
(a_0,\,p)=\left(\frac{6\,k^2\,(\alpha\gamma_2-1)}{\alpha\,(\gamma_1\gamma_2-1)},\,2\right).
\end{gathered}
\end{equation}
So,  we have  the first member of the solution expansion in the Laurent series
\begin{equation} \label{eq:1.4a}
\begin{gathered}
w\simeq\frac{6\,k^2\,(\alpha\gamma_2-1)}{\alpha\,(\gamma_1\gamma_2-1)(z-z_0)^2}+\ldots
\end{gathered}
\end{equation}

At the next step of investigation we have to find the Fuchs indices \cite{Ablowitz01, Ablowitz02}. For this purpose we substitute
\begin{equation} \label{eq:1.4b}
\begin{gathered}
w\simeq\frac{6\,k^2\,(\alpha\gamma_2+1)}{\alpha\,(\gamma_1\gamma_2+1)\,z^2}+a_jz^{j-2}
\end{gathered}
\end{equation}
into \eqref{eq:1.3c} again and equate the expressions at the first order of $a_j$. We obtain the following Fuchs indices for the~solution of Eq.\eqref{eq:1.3b}:
\begin{equation} \label{eq:1.5}
\begin{gathered}
j_1=-1, \quad j_2=6, \quad j_{3,4}=\frac{5}{2}\pm\frac{1}{2}\sqrt{\frac{24\alpha^2\gamma_2+\alpha\gamma_1\gamma_2-49\alpha+24\gamma_1}{\alpha(\gamma_1\gamma_2-1)}}.
\end{gathered}
\end{equation}
In order for Eq.\eqref{eq:1.3b} to  pass the Painlev\'e test we have to obtain the integer values for the Fuchs indices. In this case we introduce the following equality:
\begin{equation} \label{eq:1.5a}
\frac{24\alpha^2\gamma_2+\alpha\gamma_1\gamma_2-49\alpha+24\gamma_1}{\alpha(\gamma_1\gamma_2-1)}=(2m+1)^2,
\end{equation}
where $m=0,1,2,\ldots$

For $m=0$ we get $\alpha=\frac{1\pm\sqrt{1-\gamma_1\gamma_2}}{\gamma_2}$ and $j=-1,\,2,\,3,\,6$. Thus the coefficients $a_2$, $a_3$ and $a_6$ should be arbitrary values. To check this we substitute the expression
\begin{equation} \label{eq:1.5b}
\begin{gathered}
w\simeq\frac{6\,k^2\,(\alpha\gamma_2-1)}{\alpha\,(\gamma_1\gamma_2-1)(z-z_0)^2}+\frac{a_1}{z-z_0}+a_2+a_3\,(z-z_0)+a_4\,(z-z_0)^2+\\a_5\,(z-z_0)^3+a_6\,(z-z_0)^4
\end{gathered}
\end{equation}
into equation \eqref{eq:1.3b} and equate to zero the expressions at the same degrees of $(z-z_0)$. We obtain that the coefficients $a_2$, $a_3$ and $a_6$ can not be chosen arbitrary. So, in the case of $m=0$ the system does not possess the Painlev\'e property.

In the case  of $m=1$ we have $\alpha=\frac{\gamma_1\gamma_2+5\pm\sqrt{\gamma_1^2\gamma_2^2-26\gamma_1\gamma_2+25}}{6\gamma_2}$ and the Fuchs indices $j=-1,\,1,\,4,\,6$. So, the coefficients $a_1,\,a_4,\,a_6$ should be arbitrary values, but it is not the case.  We obtain that $a_1$ is an  arbitrary value, but the coefficients $a_4$ and $a_6$ can not be chosen arbitrary. So, in this case system \eqref{eq:1.3} does not possess the Painlev\'e property.

In the case of $m=2$ we have $\alpha=\frac{1}{\gamma_2},\gamma_1$ and the following Fuchs indices $j=-1,\,0,\,5,\,6$. Since $j=0$ then $a_0$ must be an arbitrary value, but it is not the case because $a_0=0$ or $a_0=\frac{6k^2}{\gamma_1}$. Thus, in the case of $m=2$ system \eqref{eq:1.3} does not pass the Painlev\'e test too.

In general case  Fuchs indices \eqref{eq:1.5}  are not integer and system \eqref{eq:1.3} does not pass the Painlev\'e test. But as $j=6$ is an integer value then the sixth coefficients $a_6$ in the~Laurent~series~\eqref{eq:1.4a} can be chosen arbitrary at special relations on the parameters of equations~\eqref{eq:1.3}. This conditions are presented in Table \ref{tabl1}.

\begin{table}[!htb]
  \footnotesize
  \caption{Relations between the parameters under which the coefficient $a_6$ can be chosen arbitrary}\label{tabl1}
  \begin{center}
    \begin{tabular}{|c|c|c|c|}
        \hline
$n$&$\alpha$&$k$&$C_0$ \\ \hline
$1$&$1$&$\pm i$&$0$ \\ \hline
$2$&$1$&$\pm 1$&$0$ \\ \hline
$3$&$1$&$\pm \frac{1}{\sqrt{6}}$&$\pm5\,k^2$ \\ \hline
$4$&$1$&$\pm \frac{1}{\sqrt{6}}\,i$&$\pm5\,k^2$ \\ \hline
$5$&$\frac{1-\gamma_1}{\gamma_2-1}$&$\pm \sqrt{\gamma_1-1}$&$0$ \\ \hline
$6$&$\frac{1-\gamma_1}{\gamma_2-1}$&$\pm \sqrt{\gamma_1-1}\, i$&$0$ \\ \hline
$7$&$\frac{1-\gamma_1}{\gamma_2-1}$&$\pm \sqrt{\frac{\gamma_1-1}{6}}$&$\pm5\,k^2$ \\ \hline
$8$&$\frac{1-\gamma_1}{\gamma_2-1}$&$\pm \sqrt{\frac{\gamma_1-1}{6}}\,i$&$\pm5\,k^2$ \\ \hline
    \end{tabular}
  \end{center}
\end{table}

\section{Traveling wave exact solutions for the Lotka--Volterra competition system}
Let us find the traveling wave exact solutions for the Lotka--Volterra competition  system. For this purpose we will use the values of parameters from Table~\ref{tabl1}.

Using traveling wave variables~\eqref{eq:1.4} we reduce system~\eqref{eq:1.3} to  system of nonlinear ordinary differential equations~\eqref{eq:1.3a}.
At the present time there are many methods for finding exact solutions of nonlinear differential equations. The most useful of them are the tanh-expansion method \cite{Hereman01, Parkes01, Biswas01}, the simplest equation method \cite{Kudr90, Kudr05aa, Vitanov01, Vitanov02}, the $G^{'}/G$-expansion method \cite{Wang, Kudr10k} and others~\cite{Vitanov03, Biswas02, PolyaninZhurov2014}. In the present work we use the method of $Q$-functions \cite{Kudr12a} (the~Kudryashov method). This method has some advantages that were discussed in recent papers \cite{KudrAMC13, KudrZakh2014, KudrZakh2014-Chaos, KudrZakh2014-MMAS}.

It was found that both equations in system~\eqref{eq:1.3a} have the second order pole, therefore we seek solutions of this system in the following form:
\begin{equation} \label{eq:1.7}
\begin{gathered}
y(z)=A_0+A_1\,Q(z)+A_2\,Q^2(z), \\
w(z)=B_0+B_1\,Q(z)+B_2\,Q^2(z),
\end{gathered}
\end{equation}
where
\begin{equation} \label{eq:1.7a}
Q(z)=\frac{1}{1+e^{-z-z_0}}
\end{equation}
is the logistic function and $z_0$ is an arbitrary constant.

It is easy to show that the function $Q(z)$ is a solution of the equation
\begin{equation} \label{eq:1.7b}
Q_z=Q-Q^2.
\end{equation}
Equation \eqref{eq:1.7b} allows us to obtain $y_z$, $y_{zz}$, $w_z$ and $w_{zz}$ using polynomials of $Q$.

Substituting $y$, $w$, $y_z$, $y_{zz}$, $w_z$ and $w_{zz}$ expressed via $Q$ into~\eqref{eq:1.3a} and equating to zero the~expressions at the same degrees of $Q$ we obtain the coefficients $A_0,\,A_1,\,A_2,\,B_0,\,B_1$ and $B_2$.

We found that it is possible to obtain exact traveling wave solutions only in the cases
\begin{equation} \label{eq:3.1}
\alpha=1,\,\frac{1-\gamma_1}{\gamma_2-1}.
\end{equation}

According to the $Q$-function method we obtain the following solutions in the case of $\alpha=1$:
\begin{equation} \label{eq:3.2}
\begin{gathered}
z=k\,x-C_0\,t,\,\,\, C^{(1)}_0=0,\\
k_{1,2}=\pm i,\,\,\,k_{3,4}=\pm 1,
\end{gathered}
\end{equation}
or
\begin{equation} \label{eq:3.3}
\begin{gathered}
z=k\,x-C_0\,t,\,\,\, C^{(2)}_0=-5\,k^2,\,\,\,\,C^{(3)}_0=5\,k^2, \\
k_{5,6}=\pm \frac{1}{\sqrt{6}}\,i,\,\,\,k_{7,8}=\pm \frac{1}{\sqrt{6}}.
\end{gathered}
\end{equation}

In the case of $\alpha=\frac{1-\gamma_1}{\gamma_2-1}$ we get
\begin{equation} \label{eq:3.4}
\begin{gathered}
z=k\,x-C_0\,t,\,\,\, C^{(1)}_0=0,\\
k_{1,2}=\pm \sqrt{\gamma_1-1},\,\,\,k_{3,4}=\pm \sqrt{\gamma_1-1}\,i,
\end{gathered}
\end{equation}
or
\begin{equation} \label{eq:3.5}
\begin{gathered}
z=k\,x-C_0\,t,\,\,\, C^{(2)}_0=-5\,k^2,\,\,\,\,C^{(3)}_0=5\,k^2, \\
k_{5,6}=\pm \sqrt{\frac{\gamma_1-1}{6}}\,i,\,\,\,k_{7,8}=\pm \sqrt{\frac{\gamma_1-1}{6}}.
\end{gathered}
\end{equation}

So, we have the following solutions of system~\eqref{eq:1.3} for  $\alpha=1$:
\begin{equation} \label{eq:3.31}
\begin{gathered}
u(x,t)=\frac{(\gamma_1-1)}{(\gamma_1\gamma_2-1)}\left(1-Q(z)\right)^2,\\
v(x,t)=\frac{(\gamma_2-1)}{(\gamma_1\gamma_2-1)}\left(1-Q(z)\right)^2,\\
z=\pm \frac{1}{\sqrt{6}}\,x-\frac{5}{6}\,t.
\end{gathered}
\end{equation}

\begin{equation} \label{eq:3.32}
\begin{gathered}
u(x,t)=\frac{(\gamma_1-1)}{(\gamma_1\gamma_2-1)}\left(1-Q(z)^2\right),\\
v(x,t)=\frac{(\gamma_2-1)}{(\gamma_1\gamma_2-1)}\left(1-Q(z)^2\right),\\
z=\pm \frac{1}{\sqrt{6}}\,i\,x-\frac{5}{6}\,t.
\end{gathered}
\end{equation}

\begin{equation} \label{eq:3.33}
\begin{gathered}
u(x,t)=\frac{(\gamma_1-1)}{(\gamma_1\gamma_2-1)}\left(1-6Q(z)+6Q(z)^2\right),\\
v(x,t)=\frac{(\gamma_2-1)}{(\gamma_1\gamma_2-1)}\left(1-6Q(z)+6Q(z)^2\right),\\
z=\pm x.
\end{gathered}
\end{equation}

\begin{equation} \label{eq:3.34}
\begin{gathered}
u(x,t)=\frac{6\,(\gamma_1-1)}{(\gamma_1\gamma_2-1)}\left(Q(z)-Q(z)^2\right),\\
v(x,t)=\frac{6\,(\gamma_2-1)}{(\gamma_1\gamma_2-1)}\left(Q(z)-Q(z)^2\right),\\
z=\pm i\,x.
\end{gathered}
\end{equation}

\begin{figure}[!h]
\center{\includegraphics[scale=0.3]{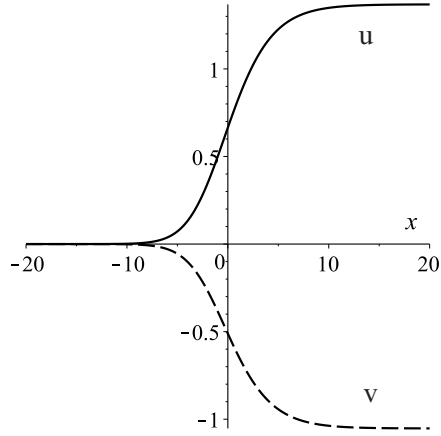} }
\caption{Solution \eqref{eq:3.31}  of system \eqref{eq:1.3} at $t=1$, $\gamma_1=0.35,\,\gamma_2=1.5$.}
\label{fig1}
\end{figure}

Solution~\eqref{eq:3.31} is shown in Fig.\ref{fig1}. This solution always takes real values. The function $u$ is positive at $\gamma_1<1,\, \gamma_2<\frac{1}{\gamma_1}$, the function $v$ is positive at $\gamma_2>1,\, \gamma_1>\frac{1}{\gamma_2}$.
The solutions are kinks which are moving with constant velocity
\begin{equation} \label{eq:3.31a}
c=\pm \frac{5}{\sqrt{6}}.
\end{equation}

Solution~\eqref{eq:3.32} always takes complex values.

\begin{figure}[!h]
\center{\includegraphics[scale=0.3]{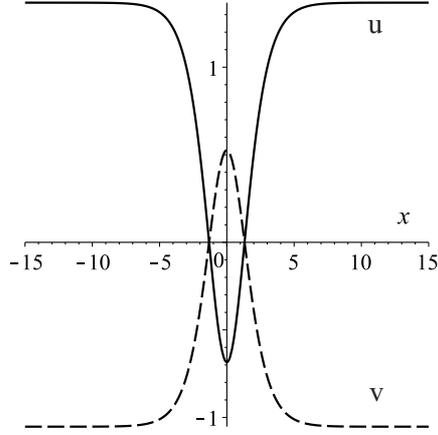} }
\caption{Solution \eqref{eq:3.33}  of system \eqref{eq:1.3} at $t=1$, $\gamma_1=0.35,\,\gamma_2=1.5$.}
\label{fig2}
\end{figure}

Fig.\ref{fig2} demonstrates stationary solution~\eqref{eq:3.33}. The solution components $u$ and $v$ are always real, but can take both positive and negative values. Stationary solution~\eqref{eq:3.34} takes complex values.

Also we have another solutions of system~\eqref{eq:1.3} for  $\alpha=\frac{1-\gamma_1}{\gamma_2-1}$:
\begin{equation} \label{eq:3.36}
\begin{gathered}
u(x,t)=Q(z)^2,\\
v(x,t)=1-Q(z)^2,\\
z=\pm \sqrt{\frac{1-\gamma_1}{6}}\,x-\frac{5\,(\gamma_1-1)}{6}\,t.
\end{gathered}
\end{equation}

\begin{equation} \label{eq:3.35}
\begin{gathered}
u(x,t)=2Q(z)-Q(z)^2,\\
v(x,t)=1-2Q(z)+Q(z)^2,\\
z=\pm \sqrt{\frac{1-\gamma_1}{6}}\,i\,x-\frac{5\,(\gamma_1-1)}{6}\,t.
\end{gathered}
\end{equation}

\begin{equation} \label{eq:3.38}
\begin{gathered}
u(x,t)=1-6Q(z)+6Q(z)^2,\\
v(x,t)=6Q(z)-6Q(z)^2,\\
z=\pm \sqrt{{1-\gamma_1}}\,x.
\end{gathered}
\end{equation}

\begin{equation} \label{eq:3.37}
\begin{gathered}
u(x,t)=6Q(z)-6Q(z)^2,\\
v(x,t)=1-6Q(z)+6Q(z)^2,\\
z=\pm \sqrt{{1-\gamma_1}}\,i\,x.
\end{gathered}
\end{equation}

\begin{figure}[!h]
\center{\includegraphics[scale=0.3]{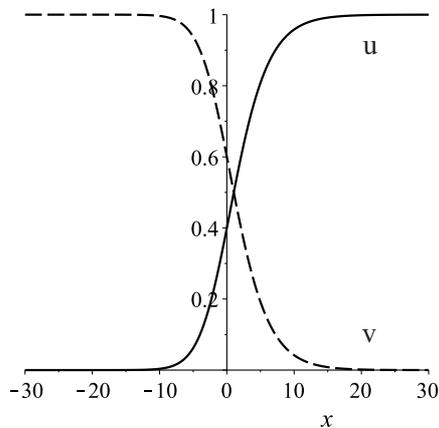} }
\caption{Solution \eqref{eq:3.36}  of system \eqref{eq:1.3} at $t=1$, $\gamma_1=0.35,\,\gamma_2=1.5$.}
\label{fig3}
\end{figure}

Fig.\ref{fig3} shows traveling wave solution~\eqref{eq:3.36}. The functions $u$ and $v$ are positive for any values of $\gamma_1,\gamma_2$. This solutions are kinks which are moving with constant velocity
\begin{equation} \label{eq:1.25a}
c=\pm 5\,\sqrt{\frac{{1-\gamma_1}}{6}}.
\end{equation} Solution~\eqref{eq:3.35} takes only complex values.

\begin{figure}[!h]
\center{\includegraphics[scale=0.3]{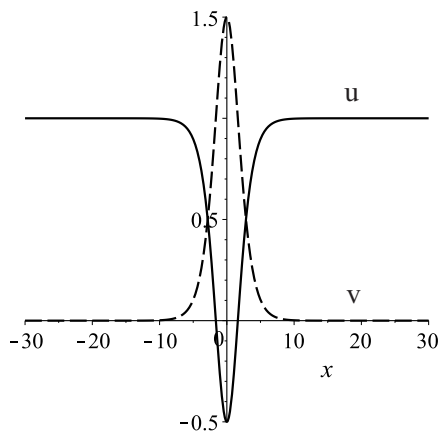} }
\caption{Solution \eqref{eq:3.38}  of system \eqref{eq:1.3} at $t=1$, $\gamma_1=0.35,\,\gamma_2=1.5$.}
\label{fig4}
\end{figure}

Stationary solution~\eqref{eq:3.38} is illustrated in Fig.\ref{fig4}. The solution components are real-valued, but can take both positive and negative values for any $\gamma_1,\gamma_2$. Solution~\eqref{eq:3.37} are complex for all values of the parameters.

We note that using the identity $Q(z)=1-Q(-z)$ one can obtain another form of presentation for obtained solutions \eqref{eq:3.31}--\eqref{eq:3.34} and \eqref{eq:3.36}--\eqref{eq:3.37}.

It should be noted that in the obtained solutions the functions $u$ and $v$ are related to each other:
\begin{equation} \label{eq:3.40}
\begin{gathered}
v(x,t)=\left(\frac{\gamma_2-1}{\gamma_1-1}\right)u(x,t),\quad  \alpha=1,
\end{gathered}
\end{equation}
and
\begin{equation} \label{eq:3.41}
\begin{gathered}
v(x,t)=1-u(x,t),\quad  \alpha=\frac{1-\gamma_1}{\gamma_2-1}.
\end{gathered}
\end{equation}
This is an obvious fact, because we seek traveling wave solutions $u$ and $v$ via the same function~$Q$.

\section{Periodic exact solutions for the Lotka--Volterra competition  system}
Let us find some periodic solutions for system of equations \eqref{eq:1.3}.
For this we use traveling wave variables
\begin{equation} \label{eq:4.0a}
\begin{gathered}
u(x,t)=y(z), \quad v(x,t)=w(z), \quad z=x-C_0\,t,
\end{gathered}
\end{equation}
and obtain the following system:
\begin{equation} \label{eq:4.0}
\begin{gathered}
{y_{zz}}+{C_0}\,y_z +y\left(1-y-\gamma_1w \right)=0,\\
w_{zz}+{C_0}\,w_z +\alpha w \left(1-w-\gamma_2y\right)=0.
\end{gathered}
\end{equation}

We look for periodic solutions expressed in terms of the Weierstrass elliptic function.
The  Weierstrass elliptic function $\wp(z)$ satisfies the following equation:
\begin{equation} \label{eq:4.1}
\begin{gathered}
\wp_{z}^{2}=4\wp^3-g_2\wp-g_3,
\end{gathered}
\end{equation}
where $g_2,\,g_3$ are the invariants.
From \eqref{eq:4.1} we get
\begin{equation} \label{eq:4.1b}
\begin{gathered}
\wp_{zz}=6\wp^2-\frac{g_2}{2}.
\end{gathered}
\end{equation}

As system~\eqref{eq:1.3} has the second order pole solution and the function $\wp(z)$ has the second order pole then we can find solutions of system \eqref{eq:1.3} in the form
\begin{equation} \label{eq:4.2}
\begin{gathered}
y(z)=A_0+A_1\,\wp(z),\\
w(z)=B_0+B_1\,\wp(z).
\end{gathered}
\end{equation}

Substitute~\eqref{eq:4.2} into~\eqref{eq:4.0} and take into account expressions~\eqref{eq:4.1} and \eqref{eq:4.1b} for the derivatives of $\wp(z)$. So, we have a polynomial for the function $\wp(z)$. After equating to zero the expressions at the same degrees of $\wp(z)$ we obtain the values of coefficients $A_0,\,A_1,\,B_0,\,B_1$.

We found only stationary solutions ($C_0=0;\, z=x$) at the following relations between the parameters:
\begin{equation} \label{eq:4.3a}
\begin{gathered}
\alpha=1, \quad g_2=\frac{1}{12},
\end{gathered}
\end{equation}
or
\begin{equation} \label{eq:4.3b}
\begin{gathered}
\alpha=\frac{1-\gamma_1}{\gamma_2-1}, \quad g_2=\frac{1}{12}(\gamma_1-1)^2.
\end{gathered}
\end{equation}

\begin{figure}[!h]
\center{\includegraphics[scale=0.3]{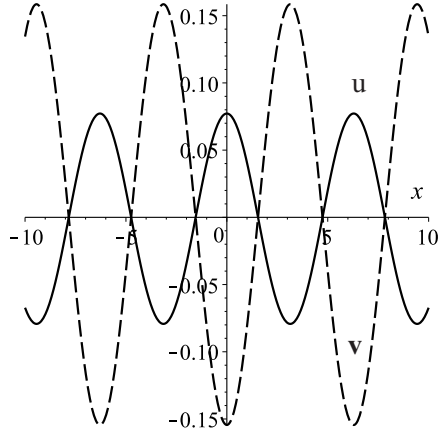} }
\caption{Periodic solution \eqref{eq:4.4a}  of system \eqref{eq:1.3} at $\gamma_1=0.75,\, \gamma_2=1.5,\,g_2=\frac{1}{12},\, g_3=\frac{1}{218}$.}
\label{fig5}
\end{figure}

In the first case we have a solution of system \eqref{eq:1.3} in the form
\begin{equation} \label{eq:4.4a}
\begin{gathered}
\begin{split}
&u(x)=\frac{6(\gamma_1-1)}{\gamma_1\gamma_2-1}\,\wp(x;g_2,g_3)+\frac{\gamma_1-1}{2(\gamma_1\gamma_2-1)}, \\
&v(x)=\frac{6(\gamma_2-1)}{\gamma_1\gamma_2-1}\,\wp(x;g_2,g_3)+\frac{\gamma_2-1}{2(\gamma_1\gamma_2-1)}.
\end{split}
\end{gathered}
\end{equation}
For real-valued solutions the invariant $g_3$ must satisfy $g_2^3-27g_3^2=\frac{1}{12^3}-27g_3^2>0$.
Fig.\ref{fig5} demonstrates periodic solution \eqref{eq:4.4a}. One can see that this solution is similar to a cnoidal wave. The points of the maximum density of the first population correspond to the points of the minimal density of another population. And there are points at which the densities of both competitive species coincide with each other.

The second case corresponds to the following solution of system \eqref{eq:1.3}:
\begin{equation} \label{eq:4.4b}
\begin{gathered}
\begin{split}
&u(x)=\frac{-6}{\gamma_1-1}\,\wp(x;g_2,g_3)+\frac{1}{2}, \\
&v(x)=\frac{6}{\gamma_1-1}\,\wp(x;g_2,g_3)+\frac{1}{2}.
\end{split}
\end{gathered}
\end{equation}

\begin{figure}[!h]
\center{\includegraphics[scale=0.3]{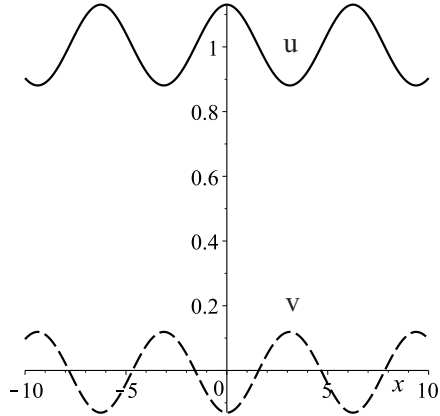} }
\caption{Periodic solution \eqref{eq:4.4b}  of system \eqref{eq:1.3} at $\gamma_1=\frac{1}{10},\, \gamma_2=1.5,\, g_2=\frac{27}{400},\,g_3=\frac{3}{1000}$.}
\label{fig6}
\end{figure}

For real-valued solution the invariant $g_3$ must satisfy $g_2^3-27g_3^2=\frac{(\gamma_1-1)^6}{12^3}-27g_3^2>0$. Periodic solution \eqref{eq:4.4b} is shown in Fig.\ref{fig6} and it is similar to a cnoidal wave too. Similarly, the maximum density of the first population correspond to the minimal density of another population. But there are no points at which the densities of competitive species equivalent to each other.

\section{Conclusion}
In this work, the system of equations describing the Lotka--Volterra competition model with diffusion was considered.

It was shown that the system does not possess the Painlev\'e property and consequently it is not integrable by the inverse scattering transform.  However we found that at some correlation between the parameters of system \eqref{eq:1.3} there are two arbitrary constants in the solution expansion in the Laurent series. By means of the Logistic-function method we obtained traveling wave exact solutions for this correlations between $\alpha,\,\gamma_1,\,\gamma_2$ . There are also standing wave solutions among them. Periodic exact solutions expressed in terms of the Weierstrass elliptic function were also obtained.

It should be noted that the obtained solutions are not presented in the handbook about solutions of nonlinear differential equations by Polyanin and Zaitsev \cite{Polyanin2011}. So, we can assume that they are new.
We also hope that these exact solutions can be used as quasi-exact solutions \cite{Kudr10m,Kudr10n} for system \eqref{eq:1.3} and \eqref{eq:1.1}. It is an interesting question which should be investigated in future.

\section*{Acknowledgements}
This work was supported by Russian Science Foundation, project to support research carried out by individual groups No.~14-11-00258.

\end{document}